# Transposition of Notations in Just Intonation

**Dr David Ryan, Edinburgh, UK**
**Draft 04, December 2016**

## 1) Abstract

A notation system was previously presented which can notate any rational frequency in free Just Intonation. Transposition of music is carried out by multiplying each member of a set of frequencies by a single frequency. Transposition of JI notations up by a fixed amount requires multiplication to be defined for any two notations. Transposition down requires inversion to be defined for any notation, which allows division to also be defined for any two notations. Each notation splits into four components which in decreasing size order are: octave, diatonic scale note, sharps or flats, rational comma adjustment. Multiplication can be defined for each of the four notation components. Since rational number multiplication is commutative, this leads to a definition of multiplication for frequencies and thus notations. Examples of notation inversion and multiplication are given. Examples of transposing melodies are given. These are checked for accuracy using the rational numbers which each notation represents. Calculation shortcuts are considered which make notation operations quicker to carry out by hand. A question regarding whether rational commas should be extended from 5–rough rational numbers to all rational numbers is considered; this would greatly simplify notation multiplication. This approach is rejected since it leads to confusion about octave number. The four component notation system is recommended instead. Extensions to computer notation systems and stave representations are briefly mentioned.

## 2) Introduction and Literature Review

Just Intonation (JI) is the theory of musical tuning in which relative frequencies are tuned to small whole–numbered ratios (such as octaves, perfect fifths, major thirds) in order to produce the most pleasant and consonant harmony. For a history of tuning and temperament the reader is referred to Fauvel, Flood & Wilson (2006), Partch (1974), Haluska (2004) and Sethares (2005). For more information about JI itself, see Doty (2002), Downes (2008) and Sabat (2009).

Two aspects of JI have been investigated in previous papers: the mathematical analysis of JI chords and scales (Ryan 2016a) and the development of a comprehensive notation for free–JI which required every prime number to be allocated a musical prime comma by algorithm (Ryan 2016b). Knowledge of this latter paper and its notation system will be assumed as the starting point here.

Other JI notation systems generally have at least one of two flaws: either they only notate a finite number of primes, or the notation becomes too long for large primes. For example the extended Helmholtz–Ellis system (Sabat 2005) introduces two new accidentals for each higher prime (5 and above), and has to stop after a finite number of primes, currently accidentals are defined up to prime 61. Alternatively, vector notation (Monzo 2016, Xenharmonic 2016) describes JI frequencies via the prime exponent vector; however this approach requires unwritably long vectors for higher $p$–limits and therefore fails to be a usable system for free–JI.

In contrast, the previously developed free–JI notation system (Ryan 2016b) solves both of these problems by encoding the higher prime information into a small (usually microtonal) frequency shift, the 'rational comma' $[x/y]$, with the approximate overall frequency determined by a Pythagorean part $L_z$ with L a 3–limit pitch class and $z$ an octave number. In this paper, the resulting notations of the form $L[x/y]_z$ will be examined, with the particular application in mind of how to transpose music written using these notations.

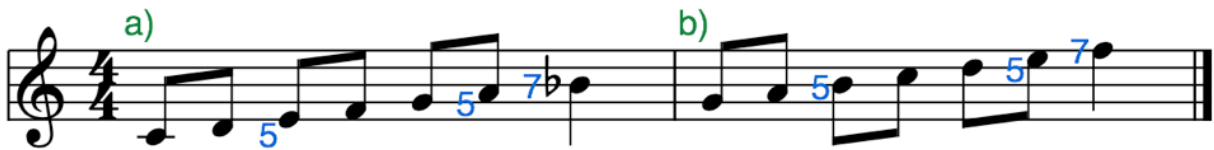

Figure 1: (a) Short melody made using JI modified diatonic scale from $C_4$ to $Bb[7]_4$, (b) a transposed repetition. Stave is 3–limit, rational commas in blue are used as accidentals to add back in the higher prime information.

The aim is to be able to calculate transpositions of JI notations. Transposing music upwards by an interval is equivalent to multiplying all the music frequencies by the transposing frequency which represents that interval. Transposing music downwards is equivalent to dividing by the transposing frequency, or to multiplication by the inverse (reciprocal) of the frequency. As a motivating example, suppose the short seven–note JI melody in Figure 1a is to be transposed up a perfect fifth to obtain the melody in Figure 1b. These two repetitions use a total of ten pitch classes, which are illustrated in Figure 2 using part of a tone lattice which is three–dimensional for these 7–limit pitch classes.

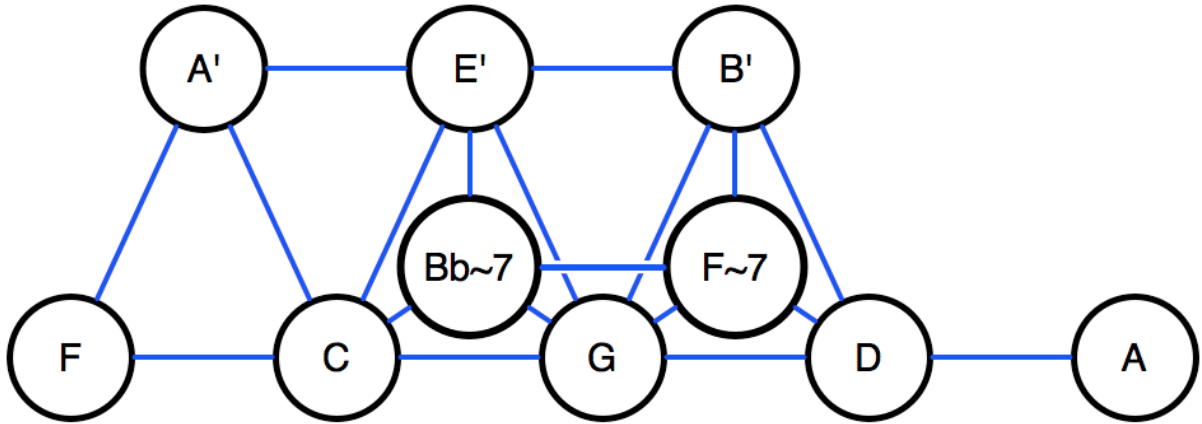

Figure 2: Illustration of pitch classes from melody above on a pitch class tone lattice.
Some shorthand used, e.g. A' = A[5] and Bb~7 = Bb[7]

The JI notation for this melody can be given as a sequence ($C_4$, $D_4$, $E'_4$, $F_4$, $G_4$, $A'_4$, $Bb[7]_4$). Some shorthand is used to reduce writing time, in particular a prime ' to represent the comma [5] = 80/81, and ~7 for a comma [7] = 63/64. See Ryan (2016b) for definitions of prime comma fractions for all prime numbers 5 and above.

The standard assignment of the identity fraction 1/1 is to the notation $C_4$. Given this definition, fractions can be calculated for the whole melody, and these are (1/1, 9/8, 5/4, 4/3, 3/2, 5/3, 7/4). Now 3/2 is a perfect fifth, and transposing the melody up by a perfect fifth is equivalent to multiplying every frequency in it by 3/2 = $G_4$ which gives the result (3/2, 27/16, 15/8, 2/1, 9/4, 5/2, 21/8). These new fractions also have notations, which are ($G_4$, $A_4$, $B'_4$, $C_5$, $D_5$, $E'_5$, $F[7]_5$).

The transposition of the melody can be represented as two different multiplications of a vector (component–wise) by a scalar:

- (1/1, 9/8, 5/4, 4/3, 3/2, 5/3, 7/4) × 3/2 = (3/2, 27/16, 15/8, 2/1, 9/4, 5/2, 21/8)
- ($C_4$, $D_4$, $E'_4$, $F_4$, $G_4$, $A'_4$, $Bb[7]_4$) × $G_4$ = ($G_4$, $A_4$, $B'_4$, $C_5$, $D_5$, $E'_5$, $F[7]_5$)

The latter representation in terms of notations gives seven facts which must be true about notation multiplication:

1. $C_4 \times G_4 = G_4$
2. $D_4 \times G_4 = A_4$
3. $E'_4 \times G_4 = B'_4$
4. $F_4 \times G_4 = C_5$
5. $G_4 \times G_4 = D_5$
6. $A'_4 \times G_4 = E'_5$
7. $Bb[7]_4 \times G_4 = F[7]_5$

These equations are just multiplications of fractions being given new names. Each multiplication on the left hand side (L.H.S.) is commutative, which means the order could be switched without affecting the result. Notation multiplication is commutative because fraction multiplication is commutative, i.e. multiplication of two numbers is the same in either order. This property makes it possible to separate out different aspects of the notation in order to multiply them separately (in any order) and then recombine. Some further observations on these transpositions/multiplications:

- $C_4 = 1/1$ is the identity element of the notation multiplication. It did not change $G_4$ when multiplied with it in line (1) above. Multiplying any notation by the identity ($C_4$) returns the same notation as a result.
- Sometimes notations in octave 4 when multiplied stay in octave 4, but sometimes they move up to octave 5, e.g. compare (2) above with (4-7).
- Higher prime information is well behaved since it is preserved during multiplication; the prime commas [5] and [7] are always present on the L.H.S. in the same quantity as on the R.H.S. ([5] = ')
- $F_4$ and $G_4$ are nearly opposite to each other since their product gives $C_5 = 2/1$. Hence by reducing the octave number of either $F_4$ or $G_4$ by 1, opposites would be obtained. This means $F_3$ and $G_4$ are opposites (in mathematical terms, inverses) of each other, as are $G_3$ and $F_4$.

In fact, notation multiplication is closed (multiplying any two notations gives another notation), has an identity ($C_4$), inverses (e.g. $D_4$ and $Bb_3$), is associative (bracketing of multiplication is irrelevant) and commutative (order of multiplication is irrelevant). Mathematically speaking, this means that notations form a group under multiplication, which is to be expected, since the underlying set of rational numbers (JI frequency fractions) form a commutative group under multiplication.

## 3) Just Intonation general notation and its alternative forms

The general form of a JI notation is $L[x/y]_z$ where:
- L contains one diatonic scale note label from (C, D, E, F, G, A, B)
- L may also have one or more sharps (#), one or more flats (b) or no sharps or flats
- $z$ is an octave number centered on 4 (since $C_4 = 1/1$)
- $[x/y]$ is a rational comma containing a microtonal adjustment for each higher prime (5 and above) in the prime factorisation of the frequency being notated. This means that $x$ and $y$ are 5–rough numbers, or numbers which only have prime factors 5 and above; these start 1, 5, 7, 11, 13… and are all the numbers which equal 1 or 5 mod 6.

If the comma is [1/1] then it can be omitted, and the notation written as $L_z$ which give the set of 3–limit (Pythagorean) JI notations. There are some alternative forms to consider for the general notation:

1. $L[x/y]_z$
2. $L_z[x/y]$
3. $Lz[x/y]$ or $L(-z)[x/y]$
4. $L[x/y]z$ or $L[x/y](-z)$

These have been presented in order of preference, based on the following considerations. Firstly, a negative $z$ could be mistaken for subtraction if it is not subscripted or bracketed. For example, $C_{-2}$ is clearly six octaves below $C_4$, however C–2 is much less clear in comparison with C4. Secondly, a positive octave number $z$ could look like multiplication if it appears at the end without being subscripted. For example, E4[5] is more clear than E[5]4.

Therefore, if rich text is available, subscripting the octave number is recommended, and $L[x/y]_z$ is preferred. Otherwise, the form L$z$[x/y] is recommended, e.g. for computer keyboard input or ASCII display output, with extra brackets round the octave number if it is negative to avoid confusion with subtraction. When selecting a notation style, it is worth considering if $L_z$ or $L[x/y]$ are a more useful grouping to display together; $L_z$ is the Pythagorean approximation, and $L[x/y]$ is the pitch class.

## 4) Notation operations and components

Musical transposition requires three operations: how to find the inverse of a notation, how to multiply two notations, and how to divide one notation by another. These three operations can be written as:

- Inversion: $(L[u/v]_w)^{-1}$
- Multiplication: $L[u/v]_w \times M[x/y]_z$
- Division: $L[u/v]_w / M[x/y]_z = L[u/v]_w \times (M[x/y]_z)^{-1}$

The third operation (division) is straightforward since division by a notation is the same as multiplication by its inverse, in either order since multiplication of fractions is commutative. Hence the first two operations of inversion and multiplication are all that need to be defined.

In fact, each notation can be separated into four components which can be considered as separate fractions multiplied together. This enables notation multiplication to be defined by how the components multiply together. Here are the four components:

- $L[x/y]_z = C_z \times N_4 \times S_k \times [x/y]$

These are: an octave shift $C_z$ by an integer $z$ number of octaves, a Pythagorean diatonic scale note $N_4$ with seven choices from the seven note labels; a 'sharp number' $S_k$ representing a shift by a whole number of sharps or flats; a microtonal rational comma $[x/y]$. These have been arranged in descending order of the usual size of each component. Examples of what values each component can take are given in the following four tables, with the component choices given in notation, fraction and cent forms:

**Table 1: Options for octave shifts $C_z$ with octave number $z$ any integer**

| $C_z$ | … | $C_{-1}$ | $C_0$ | $C_1$ | $C_2$ | $C_3$ | $C_4$ | $C_5$ | $C_6$ | $C_7$ | $C_8$ | $C_9$ | … | $C_z$ |
|---|---|---|---|---|---|---|---|---|---|---|---|---|---|---|
| **Fraction** | … | 1/32 | 1/16 | 1/8 | 1/4 | 1/2 | 1/1 | 2/1 | 4/1 | 8/1 | 16/1 | 32/1 | … | $2^{z-4}$ |
| **Cents** | … | -6000 | -4800 | -3600 | -2400 | -1200 | 0 | 1200 | 2400 | 3600 | 4800 | 6000 | … | |

**Table 2: Options for the seven Pythagorean diatonic scale notes $N_4$ in octave 4**

| $N_4$ | $C_4$ | $D_4$ | $E_4$ | $F_4$ | $G_4$ | $A_4$ | $B_4$ |
|---|---|---|---|---|---|---|---|
| **Fraction** | 1/1 | 9/8 | 81/64 | 4/3 | 3/2 | 27/16 | 243/128 |
| **Cents** | 0 | 203.91 | 407.82 | 498.04 | 701.96 | 905.87 | 1109.78 |

**Table 3: Options for $S_k$, the 'sharp number' representing sharps for $k > 0$ or flats for $k < 0$**

| $S_k$ | … | $S_{-5}$ | $S_{-4}$ | $S_{-3}$ | $S_{-2}$ | $S_{-1}$ | $S_0$ | $S_1$ | $S_2$ | $S_3$ | $S_4$ | $S_5$ | … |
|---|---|---|---|---|---|---|---|---|---|---|---|---|---|
| Flats or Sharps | … | bbbbb | bbbb | bbb | bb | b | $C_4$ | # | ## | ### | #### | ##### | … |
| Fraction | | | | | … | 2048/2187 | 1/1 | 2187/2048 | … | | | | |
| Cents | … | -568 | -455 | -341 | -227 | -113.685 | 0 | 113.685 | 227 | 341 | 455 | 568 | … |

**Table 4: Some options for rational comma [x/y] where x and y can be any 5–rough numbers**

| [x/y] | Empty or [1/1] | [5] [5/1] | [1/5] | [7] [7/1] | [11] | [13] | [17] | [19] | [5/7] [5][7]$^{-1}$ | [25] [5]$^2$ | [35] [5][7] |
|---|---|---|---|---|---|---|---|---|---|---|---|
| Fraction | 1/1 | 80/81 | 81/80 | 63/64 | 33/32 | 26/27 | 2176/2187 | 513/512 | 5120/5103 | 6400/6561 | 35/36 |
| Cents | 0 | -21.51 | 21.51 | -27.26 | 53.27 | -65.34 | -8.73 | 3.38 | 5.76 | -43.01 | -48.77 |

Each component might be equal to 1/1 which means it does not contribute any frequency shift; these are shaded grey in the tables above. In some cases the component can then be omitted from the notation, e.g. write $G_5$ not $G[1/1]_5$, and write nothing for $S_0$ which is no sharps or flats.

Three out of four components are easy to define multiplication for:

- $S_m \times S_n = S_{m+n}$    (count sharps and flats, which cancel each other out)
- $C_m \times C_n = C_{m+n-4}$    (count octaves, deduct 4 since $C_4 \times C_4 = C_4$)
- $[u/v] \times [x/y] = [(ux)/(vy)]$    (rational commas combine easily, $(ux)/(vy)$ may simplify)

It is straightforward to find inverses for these:

- $S_m^{-1} = S_{-m}$    (sharps and flats are inverses of each other)
- $C_m^{-1} = C_{8-m}$    (octaves form pairs of inverses, such as $C_3$ and $C_5$)
- $[x/y]^{-1} = [y/x]$    (invert a rational number by turning it upside down)

Multiplication of scale notes $N_4$ is more complicated. This may modify both sharp number and octave number, for example:

- $F_4 \times F_4 = Bb_4$
- $D_4 \times E_4 = F\#_4$
- $G_4 \times A_4 = E_5$

The result of the $N_4$ multiplication does not affect the rational comma, that component is entirely separate from the others. However, the $N_4$ result might increase or decrease sharp number by 1, and might increase octave number by 1. Also, there is no single formula for the result, so a 7–by–7 multiplication table is needed; this is presented in the next section.

## 5) Multiplication and inversion of the diatonic scale notes

The seven diatonic scale notes in octave 4 are given in Table 2 above. Since each notation corresponds to a fraction, a 7–by–7 multiplication table can be constructed. This table is presented twice: first ordered by which power of 3 is present in the fraction (Table 5), then ordered by ascending frequency (Table 6):

**Table 5: Multiplication table for notations of the form $N_4$ where N is from (F, C, G, D, A, E, B)**

| × | $F_4$ 4/3 | $C_4$ 1/1 | $G_4$ 3/2 | $D_4$ 9/8 | $A_4$ 27/16 | $E_4$ 81/64 | $B_4$ 243/128 |
|---|---|---|---|---|---|---|---|
| $F_4$ 4/3 | $Bb_4$ 16/9 | $F_4$ 4/3 | $C_5$ 2/1 | $G_4$ 3/2 | $D_5$ 9/4 | $A_4$ 27/16 | $E_5$ 81/32 |
| $C_4$ 1/1 | $F_4$ 4/3 | $C_4$ 1/1 | $G_4$ 3/2 | $D_4$ 9/8 | $A_4$ 27/16 | $E_4$ 81/64 | $B_4$ 243/128 |
| $G_4$ 3/2 | $C_5$ 2/1 | $G_4$ 3/2 | $D_5$ 9/4 | $A_4$ 27/16 | $E_5$ 81/32 | $B_4$ 243/128 | $F\#_5$ 729/256 |
| $D_4$ 9/8 | $G_4$ 3/2 | $D_4$ 9/8 | $A_4$ 27/16 | $E_4$ 81/64 | $B_4$ 243/128 | $F\#_4$ 729/512 | $C\#_5$ 2187/1024 |
| $A_4$ 27/16 | $D_5$ 9/4 | $A_4$ 27/16 | $E_5$ 81/32 | $B_4$ 243/128 | $F\#_5$ 729/256 | $C\#_5$ 2187/1024 | $G\#_5$ 6561/2048 |
| $E_4$ 81/64 | $A_4$ 27/16 | $E_4$ 81/64 | $B_4$ 243/128 | $F\#_4$ 729/512 | $C\#_5$ 2187/1024 | $G\#_4$ 6561/4096 | $D\#_5$ 19683/8192 |
| $B_4$ 243/128 | $E_5$ 81/32 | $B_4$ 243/128 | $F\#_5$ 729/256 | $C\#_5$ 2187/1024 | $G\#_5$ 6561/2048 | $D\#_5$ 19683/8192 | $A\#_5$ 59049/16384 |

In Table 5 it can be seen that multiplying notes with diatonic scale note labels (F, C, G, D, A, E, B) results in notes with the Pythagorean labels (Bb, F, C, G, D, A, E, B, F#, C#, G#, D#, A#). This series of labels is found diagonally from upper left to lower right. Each label occupies a diagonal from lower left to upper right. Results from octave 4 and octave 5 appear in mixed order, for example $D_4 \times D_4 = E_4$ and $F_4 \times B_4 = E_5$. For the sharped labels (F#, C#, G#, D#, A#) the 3–limit numbers in the fractions get quite large, between 3 and 5 decimal digits.

**Table 6: Multiplication table for notations of the form $N_4$ where N is from (C, D, E, F, G, A, B)**

| × | $C_4$ 1/1 | $D_4$ 9/8 | $E_4$ 81/64 | $F_4$ 4/3 | $G_4$ 3/2 | $A_4$ 27/16 | $B_4$ 243/128 |
|---|---|---|---|---|---|---|---|
| $C_4$ 1/1 | $C_4$ 1/1 | $D_4$ 9/8 | $E_4$ 81/64 | $F_4$ 4/3 | $G_4$ 3/2 | $A_4$ 27/16 | $B_4$ 243/128 |
| $D_4$ 9/8 | $D_4$ 9/8 | $E_4$ 81/64 | $F\#_4$ 729/512 | $G_4$ 3/2 | $A_4$ 27/16 | $B_4$ 243/128 | $C\#_5$ 2187/1024 |
| $E_4$ 81/64 | $E_4$ 81/64 | $F\#_4$ 729/512 | $G\#_4$ 6561/4096 | $A_4$ 27/16 | $B_4$ 243/128 | $C\#_5$ 2187/1024 | $D\#_5$ 19683/8192 |
| $F_4$ 4/3 | $F_4$ 4/3 | $G_4$ 3/2 | $A_4$ 27/16 | $Bb_4$ 16/9 | $C_5$ 2/1 | $D_5$ 9/4 | $E_5$ 81/32 |
| $G_4$ 3/2 | $G_4$ 3/2 | $A_4$ 27/16 | $B_4$ 243/128 | $C_5$ 2/1 | $D_5$ 9/4 | $E_5$ 81/32 | $F\#_5$ 729/256 |

| A₄ 27/16 | A₄ 27/16 | B₄ 243/128 | C#₅ 2187/1024 | D₅ 9/4 | E₅ 81/32 | F#₅ 729/256 | G#₅ 6561/2048 |
|---|---|---|---|---|---|---|---|
| B₄ 243/128 | B₄ 243/128 | C#₅ 2187/1024 | D#₅ 19683/8192 | E₅ 81/32 | F#₅ 729/256 | G#₅ 6561/2048 | A#₅ 59049/16384 |

In Table 6 the same set of multiplications are presented, but in order of ascending frequency (C, D, E, F, G, A, B). At first glance, labels are found in diagonals from lower left to upper right, however upon closer inspection these diagonals mix sharped and flatted versions of the same label, e.g. F and F#, G and G#, B and Bb, C and C#, D and D#. Octave number behaves nicely with a value of 4 in the top left of the table and 5 in the bottom right, transitioning from octave 4 to octave 5 between the B/Bb and C/C# diagonals.

**Table 7: Inversion table for notations of the form $N_4$ where N is from (C, D, E, F, G, A, B)**

| Note | $C_4$ 1/1 | $D_4$ 9/8 | $E_4$ 81/64 | $F_4$ 4/3 | $G_4$ 3/2 | $A_4$ 27/16 | $B_4$ 243/128 |
|---|---|---|---|---|---|---|---|
| Note Inverse | $C_4$ 1/1 | $Bb_3$ 8/9 | $Ab_3$ 64/81 | $G_3$ 3/4 | $F_3$ 2/3 | $Eb_3$ 16/27 | $Db_3$ 128/243 |

In Table 7 the inversion table for scale notes is given. Except for $C_4$, all of the inverse notes are between 1/2 and 1/1, and consequently are in octave 3. Four out of seven require flats to describe them. Hence inverting a notation might change the sharp number and the octave number.

Enough information has now been given to be able to invert, multiply or divide any two JI notations, and hence perform any transpositions up or down. The next three sections will give some examples, and the reader can skip ahead to section (9) if satisfied that the calculation method works.

## 6) Examples of inverting compound notations

A compound notation is one which requires at least two non–trivial components to describe it. By that definition, $C_5$, $F_4$, $C\#_4$, $C[5]_4$ are all simple notations, and their product $F\#[5]_5$ is a compound notation.

<u>Example 1</u> Invert $E[5]_6$ (which is part of a C major triad two octaves above middle–C):

$( E[5]_6 )^{-1} = ( C_6 \times E_4 \times S_0 \times [5] )^{-1}$
$\quad\quad\quad = ( C_6 \times E_4 \times [5] )^{-1}$
$\quad\quad\quad = C_6^{-1} \times E_4^{-1} \times [5]^{-1}$
$\quad\quad\quad = C_2 \times Ab_3 \times [1/5]$
$\quad\quad\quad = C_1 \times Ab_4 \times [1/5]$
$\quad\quad\quad = Ab[1/5]_1$
$\quad\quad\quad = Ab._1 \quad\quad\quad$ (a period character . is shorthand for a rational comma [1/5])

To check this is correct:

$E[5]_6 = C_6 \times E_4 \times [5] = (4/1) \times (81/64) \times (80/81) = 25920/5184 = 5/1$
$Ab._1 = C_1 \times A_4 \times S_{-1} \times [1/5] = (1/8) \times (27/16) \times (2048/2187) \times (81/80) = 4478976/22394880 = 1/5$

So since $(5/1)^{-1} = (1/5)$, these two are indeed inverses. When confidence in inversion of notations has developed, it would no longer be necessary to carry out this manual check. The numbers can get quite large, making the manual check cumbersome. A good reason to use notations is to avoid this type of manual calculation, also to provide more musical insight into the notes than fractions give. In the cases of $E'_6 = 5/1$ and $Ab._{1} = 1/5$, the former is two octaves ($C_6 = 4/1$) and a major third ($E'_4 = 5/4$) above middle–C; the latter is the same distance below middle–C.

Example 2 Invert $Fbb..[7]_3 = Fbb[7/25]_3$ (this note is a dominant seventh from Gbb.. pitch class):

$( Fbb[7/25]_3 )^{-1} = ( C_3 \times F_4 \times S_{-2} \times [7/25] )^{-1}$

$\phantom{( Fbb[7/25]_3 )^{-1}} = C_5 \times G_3 \times S_2 \times [25/7]$

$\phantom{( Fbb[7/25]_3 )^{-1}} = C_5 \times G\#\#_3 \times [25/7]$

$\phantom{( Fbb[7/25]_3 )^{-1}} = C_4 \times G\#\#_4 \times [25/7]$

$\phantom{( Fbb[7/25]_3 )^{-1}} = G\#\#[25/7]_4$

$\phantom{( Fbb[7/25]_3 )^{-1}} = G\#\#"[1/7]_4$

To check this is correct:

$Fbb..[7]_3 = C_3 \times F_4 \times S_{-2} \times [7] \times [1/5] \times [1/5]$

$\phantom{Fbb..[7]_3} = (1/2) \times (4/3) \times (2048/2187)^2 \times (63/64) \times (81/80)^2 = 3584/6075 \quad (-913.58 \text{ cents})$

$G\#\#"[1/7]_4 = G_4 \times S_2 \times [5] \times [5] \times [1/7]$

$\phantom{G\#\#"[1/7]_4} = (3/2) \times (2187/2048)^2 \times (80/81)^2 \times (64/63) = 6075/3584 \quad (+913.58 \text{ cents})$

Hence the inversion scheme set out above allows any JI notation to be inverted, whilst retaining musical insight into how the notation fits into the wider harmonic scheme, unlike simply turning a rational number upside down. To make inversion easier, Table 7 could be extended to cover more common inversions; this is left as an exercise for the reader.

## 7) Examples of transposing compound notations

Transposing a notation upwards by another notation is the same as multiplying the two notations together. Transposing a notation downwards by a notation is to divide the first by the second, equivalent to multiplying the first by the inverse of the second. Since inversion has already been described, only examples of notation multiplication need to be demonstrated:

Example 1 Multiply $F\#[5]_5$ by $Eb[1/5]_6$

$F\#[5]_5 = C_5 \times F_4 \times S_1 \times [5]$

$Eb[1/5]_6 = C_6 \times E_4 \times S_{-1} \times [1/5]$

$F\#[5]_5 \times Eb[1/5]_6 = ( C_5 \times C_6 ) \times ( F_4 \times E_4 ) \times ( S_1 \times S_{-1} ) \times ( [5] \times [1/5] )$

$\phantom{F\#[5]_5 \times Eb[1/5]_6} = C_7 \times A_4 \times S_0 \times [5/5]$

$\phantom{F\#[5]_5 \times Eb[1/5]_6} = C_7 \times A_4$

$\phantom{F\#[5]_5 \times Eb[1/5]_6} = A_7$

In this example some of the components cancel out. The sharp and flat cancel out, and the [5] prime comma cancels the [1/5] rational comma. The remaining frequency is an A, up three and a half octaves from middle–C. As a numeric check on the notation multiplication:

$$F\#[5]_5 = (2/1)(4/3)(2187/2048)(80/81) = 45/16$$
$$Eb[1/5]_6 = (4/1)(81/64)(2048/2187)(81/80) = 24/5$$
$$A_7 = (8/1)(27/16) = 27/2$$
Can easily check that $(45/16)(24/5) = 27/2$

Hence this multiplication calculation is correct.

Example 2 As a more complex example, multiply $A\#\#\#[77/13]_{-5}$ by $Bbbbbb[23/55]_9$

$$A\#\#\#[77/13]_{-5} = C_{-5} \times A_4 \times S_3 \times [77/13]$$
$$Bbbbbb[23/55]_9 = C_9 \times B_4 \times S_{-5} \times [23/55]$$
$$A\#\#\#[77/13]_{-5} \times Bbbbbb[23/55]_9 = (C_{-5} \times C_9) \times (A_4 \times B_4) \times (S_3 \times S_{-5}) \times ([77/13] \times [23/55])$$
$$= (C_{-5+9-4}) \times G\#_5 \times S_{3-5} \times [1771/715]$$
$$= C_0 \times (C_5 \times G_4 \times S_1) \times S_{-2} \times [161/65]$$
$$= (C_{0+5-4}) \times G_4 \times S_{1-2} \times [161/65]$$
$$= C_1 \times G_4 \times S_{-1} \times [161/65]$$
$$= Gb[161/65]_1$$

In this example there are cancellations in all of: sharp numbers, octave numbers, rational commas. Also, since $A_4 \times B_4$ is a G# in the next octave up, this gives a further interaction with both the sharp number and the octave number. Care is needed when evaluating complex notation transpositions like these. This example is deliberately complicated in order to demonstrate that the method handles difficult notations well. It is not anticipated that composers would frequently encounter such remote notes! Again, check the calculation is correct using the underlying fractions and their prime factorisations:

$$A\#\#\#[77/13]_{-5} = 2^{-5-4}(27/16)(2187/2048)^3(63/64)(33/32)(27/26) = 2^{-58}3^{30}7^1 11^1 13^{-1}$$
$$Bbbbbb[23/55]_9 = 2^{9-4}(243/128)(2048/2187)^5(736/729)(81/80)(32/33) = 2^{59}3^{-33}5^{-1}11^{-1}23^1$$
$$Gb[161/65]_1 = (1/8)(3/2)(2048/2187)(63/64)(736/729)(81/80)(27/26) = 2^1 3^{-3} 5^{-1} 7^1 13^{-1} 23^1$$

Can check that the first two sets of indices add up to the final set of indices

So the complex calculation is correct!

## 8) Examples of transposing whole melodies

Transposing a whole melody (or harmony) is the same as multiplying each notation in a set by the transposing notation. Two examples:

Example 1: Multiply the melody ($C_4$ $Db._4$ $Eb._4$ $Ab._3$) by the transposing note $A[13]_3 = 13/16$:

$C_4$ is the identity, so the first transposed note will be $A[13]_3$ itself

$$Db._4 \times A[13]_3 = (C_4 \times C_3) \times (D_4 \times A_4) \times (S_{-1} \times S_0) \times ([1/5] \times [13])$$
$$= C_3 \times B_4 \times S_{-1} \times [13/5]$$
$$= Bb[13/5]_3$$
$$= Bb.[13]_3$$

For $Eb._4$ it is 9/8 above $Db._4$ so the transposed notation must be 9/8 above $Bb.[13]_3$ and this can be built up by parts:

$$B_3 \times 9/8 = C\#_4$$
$$Bb_3 \times 9/8 = C_4$$
$$Bb.[13]_3 \times 9/8 = C.[13]_4$$

For Ab.$_3$ it is 3/2 below Eb.$_4$ so the transposed note will be 3/2 below C.[13]$_4$
$$C_4 \times 2/3 = F_3$$
$$C.[13]_4 \times 2/3 = F.[13]_3$$
Hence the complete transposed melody is ( A[13]$_3$ Bb.[13]$_3$ C.[13]$_4$ F.[13]$_3$ ). Some shortcuts to melody transposition can be found since the notes in the melody usually have simple harmonic relations between them, which are easy to carry out upon successive transposed notations. To check this has worked, carry out a numeric calculation on the final note:
  Melody note Ab.$_3$ is (1/2)(27/16)(2048/2187)(81/80) = 4/5
  Transposing note A[13]$_3$ is (1/2)(27/16)(26/27) = 13/16
  Transposed note F.[13]$_3$ is (1/2)(4/3)(81/80)(26/27) = 13/20
  and (4/5)(13/16) = (13/16)(16/20) = 13/20 as required.
So the calculation is correct.

Example 2: Multiply the melody (Bb$_5$ C.[17]$_6$ Ebb.[19]$_6$) by the transposing note F#[23]$_2$:

Bb$_5$ × F#[23]$_2$ = ( C$_5$ × C$_2$ ) × ( B$_4$ × F$_4$ ) × ( S$_{-1}$ × S$_1$ ) × ( [1/1] × [23/1] )
  = C$_3$ × E$_5$ × S$_0$ × [23]
  = C$_4$ × E$_4$ × [23]
  = E[23]$_4$

C.[17]$_6$ × F#[23]$_2$ = ( C$_6$ × C$_2$ ) × ( C$_4$ × F$_4$ ) × ( S$_0$ × S$_1$ ) × ( [17/5] × [23/1] )
  = C$_4$ × F$_4$ × S$_1$ × [391/5]
  = F#$_4$ × [391/5]
  = F#.[391]$_4$

Ebb.[19]$_6$ × F#[23]$_2$ = ( C$_6$ × C$_2$ ) × ( E$_4$ × F$_4$ ) × ( S$_{-2}$ × S$_1$ ) × ( [19/5] × [23/1] )
  = C$_4$ × A$_4$ × S$_{-1}$ × [437/5]
  = Ab$_4$ × [437/5]
  = Ab.[437]$_4$

So the transposed melody is ( E[23]$_4$ F#.[391]$_4$ Ab.[437]$_4$ ). There is a factor of [23] in each note, so this melody could also be expressed as ( E$_4$ F#.[17]$_4$ Ab.[19]$_4$ )[23]. Again, carry out a spot check on the final note:
  Ebb.[19]$_6$ = (4/1)(81/64)(2048/2187)$^2$(81/80)(513/512) = 608/135
  F#[23]$_2$ = (1/4)(4/3)(2187/2048)(736/729) = 23/64
  Ab.$_4$[19][23] = (1/1)(27/16)(2048/2187)(81/80)(513/512)(736/729) = 437/270
  It is then simple to check (608/135)(23/64) does indeed equal 437/270
Again, the calculation is correct.

## 9) Calculation shortcuts

These calculations of inversion and multiplication can be complex. Consider what kinds of shortcut might be possible:
- Working from the simple inverses given in Table 7 to build up inverses with different octave or sharp number:

- Since inverse of $F_4$ is $G_3$, then inverse of $F\#_4$ must be $Gb_3$
- Since inverse of $G_4$ is $F_3$, then inverse of $G_5$ must be $F_2$
- Recognising some simple multiplication shortcuts:
  - if $D_4 \times E_4 = F\#_4$ then both $Db_4 \times E_4 = F_4$ and $D_4 \times Eb_4 = F_4$
  - $Bb_4 \times E_4 = D_4$ so $Bb._4 \times E_4 = D._4$ so $Bb._4 \times E_5 = D._5$ so $Bb._4 \times E[7]_5 = D.[7]_5$

It is possible to build up some complicated multiplications/transpositions very quickly by adding components to an existing transposition. Overall there are at least three ways of carrying out these operations:

- Break everything down into numeric fractions and then recombine
- Break everything down into the four notation components and then recombine
- Build up complex inversions and transpositions from simpler ones

Any or all of these methods could be helpful to the JI composer or music theorist. They help build more complex melodies and harmonies from simpler ones, by inverting or transposing one step at a time.

Further work recommended would be to develop a computer calculation tool which can take notations as ASCII input and carry out all the operations demonstrated above. Another helpful computer tool would be to have notations represented on the stave (as in Figure 1) where the stave was 3–limit and all higher limit information was specified using rational commas as accidentals. In this situation a computer function could be given to transpose segments of melody or harmony, using the calculation techniques set out above.

## 10) Should the comma function be extended to a complete notation?

A short diversion into an alternative notation form where the rational comma function $[x/y]$ contains all the note information, not just the microtonal shifts due to higher primes, and becomes a completely multiplicative function over the rational numbers. Earlier versions of this paper used a multiplicative function as the basis of notation. However as this section develops it will become apparent why the preferred approach is now a hybrid notation such as $L[x/y]_z$.

The rational comma function $[x/y]$ is a shorthand for a set of microtonal alterations and is the product of a set of $[p]$ and $[1/q]$ for $p$ ranging over prime factorisation of $x$ (with multiplicity) and $q$ likewise ranging over that of $y$.

So far, $x$ and $y$ have only been allowed to have prime factors 5 and above, i.e. 5–rough numbers. Is there a way of extending the function $[x/y]$ to a completely multiplicative function over the rational numbers? This is a function where $[a/b][c/d] = [(ac)/(bd)]$ for any integers $a$, $b$, $c$, $d$ with $c$ and $d$ not zero.

This could be done by defining only two extra values $[2]$ and $[3]$ for the two missing primes 2 and 3. What would the best choices be?

The choice for $[2]$ should be of the form $2^a$ and be able to produce $2^b$ for $b$ any integer. There are only two choices which will do this; $a = 1$ and $a = -1$. The natural choice is $a = 1$ with a positive power of 2, so that $[2] = 2/1$ would be the octave. This corresponds to $C_5 = 2/1$ above, and then $k$ in $[2^k] = 2^k$ would function like an octave number.

The natural choice for [3] would be an expression of the form $2^a 3$ which is one of:
  (…3/16, 3/8, 3/4, 3/2, 3/1, 6/1, 12/1…)
and these are all of the G pitch class, for example $G_3$ = 3/4, $G_4$ = 3/2, etc.

In the comma function definitions (see Ryan 2016b) the comma algorithm was developed by trying to keep the numbers in the fractions small and also keep the actual fraction as close to 1/1 as possible. These criteria are somewhat incompatible, and the solution found was to multiply two suitable logarithmic measures (*LCY*, *AO*) together to obtain another measure (*CM*); subsequently, minimising *CM* is the same as carrying out a trade–off of the original criteria. Here is a comparison of the relevant measures *LCY*, *AO*, *CM* over the candidate fractions $2^a 3$:

**Table 8: Candidates fractions for [3] with measures *CY*, *LCY*, *AO* and *CM***

| [3] (fraction) | … | 3/16 | 3/8 | 3/4 | 3/2 | 3/1 | 6/1 | 12/1 | 24/1 | … |
|---|---|---|---|---|---|---|---|---|---|---|
| [3] (decimal) | … | 0.1875 | 0.375 | 0.75 | 1.5 | 3 | 6 | 12 | 24 | … |
| *Complexity* *CY* | … | 48 | 24 | 12 | 6 | 3 | 6 | 12 | 24 | … |
| $\log_2(CY)$ *LCY* | … | 5.585 | 4.585 | 3.585 | 2.585 | 1.585 | 2.585 | 3.585 | 4.585 | … |
| **Absolute Octaves** $AO = |\log_2[3]|$ | … | 2.415 | 1.415 | 0.415 | 0.585 | 1.585 | 2.585 | 3.585 | 4.585 | … |
| '*Comma*' *Measure* $CM = AO \times LCY$ | … | 13.488 | 6.488 | 1.488 | 1.512 | 2.512 | 6.682 | 12.852 | 21.022 | … |

For the higher primes 5, 7… their commas [5], [7]… were less than a semitone (100 cents) and can be considered proper commas. However, the smallest two candidate fractions for [3] are 3/4 at 0.415 octaves (498 cents) and 3/2 at 0.585 octaves (702 cents). These are too big to be called 'commas', so [3] is a fraction not a comma. Likewise, the 'comma measure' *CM* should really be called a 'fraction measure' in this context.

One minimal (optimal) value of *CM* has been found (highlighted in green) which is for the candidate fraction 3/4, and also a very nearly optimal value (highlighted in light blue) has been found for 3/2.

Choosing [3] = 3/4 minimises *CM* and produces the G closest to middle–C. This would imply that:
  $[3]^2 = 9/16 = D_3$
  $[3]^3 = 27/64 = A_2$
  $[3]^4 = 81/256 = E_2$
  Also that $[3]^{-1} = 4/3 = F_4$

Choosing [3] = 3/2 has slightly higher *CM*, but a better *LCY* value, then:
  $[3]^2 = 9/4 = D_5$
  $[3]^3 = 27/8 = A_5$
  $[3]^4 = 81/16 = E_6$
  Also that $[3]^{-1} = 2/3 = F_3$

Were [3] to be chosen as either of these options, taking powers of [3] would quickly produce notes which are scattered across different octaves. Ideally the diatonic scale in octave 4 could be represented similarly to:

$[1], [3]^2, [3]^4, [3]^{-1}, [3], [3]^3, [3]^5, [2]$

However, powers of [2] are required to get each power of [3] into octave 4, and the diatonic scale for [3] = 3/4 would actually look like:

$[1], [3]^2[2], [3]^4[2]^2, [3]^{-1}, [3][2], [3]^3[2]^2, [3]^5[2]^3, [2]$

The scale for the choice [3] = 3/2 would look similar, with [2] scattered around the scale.

So **any** choice of [3] will produce large–scale confusion of octave number in the whole notation. This property being undesirable, it is now recommended to **use the L[*x/y*]$_z$ system of notation** exclusively, and not try to base a notation system on a multiplicative function which extends a comma function by defining [2] and [3]. The author has over the last two years (Dec 2016) experimented with both notation systems, and found that a system based on a [3] definition is far too confusing due to octave jumps between most nearby notes. The L[*x/y*]$_z$ notation system is much clearer since L$_z$ gives an approximate pitch for every note; it also lends itself to direct stave annotation (see Figure 1 above) where the stave is 3–limit and all that remains is to use any commas as accidentals on the stave.

Nonetheless, by choosing [2] = 2/1 and [3] = 3/4, it is possible to define any rational JI frequency in terms of a completely multiplicative function alone. This makes transposition completely straightforward. It simplifies notation algebra at the price of confusing octave numbers. Such a system might be suitable as an internal language for computers to use when describing JI notes, and always converting to a stave display or L[*x/y*]$_z$ type notation when presented to the end user.

## 11)    Conclusions

Any free–JI pitch classes can be presented in the format L[*x/y*] and any free–JI frequencies can be presented in rich text as L[*x/y*]$_z$ and in ASCII plain text as L*z*[*x/y*]. Some shorthands were recommended, in particular a prime ' for the comma [5], a period . for the comma [1/5], a shorthand L~*x*_*y* for pitch class L[*x/y*].

Transposition of notations up or down can be described by the three operations of multiplying, dividing and inverting notations. A notation L[*x/y*]$_z$ for any frequency splits into four components
$C_z \times N_4 \times S_k \times [x/y]$ which are: octave shift, diatonic scale note, sharps or flat shift, rational comma adjustment.

Multiplication for any two notations is defined by how the four components interact. To multiply two general frequencies, split each into four components, multiply the components, then recombine to find the final notation. Inversion of notations map between the underlying rational number and its reciprocal. Again, inversion can be defined by how the four components invert. Division of two notations can be defined as the first notation multiplied by the inversion of the second notation.

Although splitting frequencies into four components to carry out these operations can be cumbersome, calculation shortcuts are available, mainly based on taking an equation of simple notations and performing successive operations to build up a more complex equation. These algebraic manipulations are valid since they represent multiplying the underlying frequencies on both sides by the same frequency.

It is possible to extend the rational comma function from 5–rough numbers to all rational numbers. In this situation, [2] would describe large–magnitude octave shifts, [3] would describe medium–magnitude scale–note shifts, and [5], [7]… would describe small–magnitude microtonal shifts for higher primes. One advantage would be to make operations of multiplication, division and inversion of notations very simple to carry out, they would be the same as those on the rational numbers inside the comma brackets. However, the definition of [3] is problematic because any definition makes the octave numbers jump about for a diatonic scale or for most melodies using nearby notes. Subsequently, a notation system based on multiplicative functions is difficult for humans to use or understand. Experience has shown that a notation system such as $L[x/y]_z$ is much easier to use.

It is therefore recommended to let the rational comma remain 5–rough and denote only the higher prime information, and for the Pythagorean component to be notated by $L_z$ which concisely contains all relevant information, and provides easy access to the pitch class and octave height.

Computerised extensions were considered which would allow computers to: use $L[x/y]_z$ type notation systems as an internal language; calculate notation transpositions and inversions on behalf of the user; visually present the notation system on a 3–limit stave using commas as accidentals where necessary; enable carrying out transposition and inversion operations upon blocks of notes using visual on–screen commands; allow existing stave notation software to be converted for free–JI composition. Options for visual presentation of free–JI staves is expected to be the subject of the next paper.

## 12) Nomenclature and Abbreviations

*AO*  The *AbsoluteOctaves* function, equivalent to $|\log_2(x/y)|$ for a fraction $x/y$

ASCII  Americal Standard Code for Information Interchange. Contains a set of characters which can be entered via a standard computer keyboard into a computer plain text file. Does not incorporate rich text effects such as bolding and subscripting.

b  Flat character in ASCII, equivalent to fraction 2048/2187
 (B is a note label, b is a flat, so notation is case sensitive!)

Cent  A (non–JI) interval which is one hundredth of a semitone, of size $2^{1/1200} = 1.000578…$

Comma  An interval which is smaller than a semitone

Compound  A compound notation requires at least 2 non–trivial components to describe it

*CM*  The *CommaMeasure* function, equal to *AO* multiplied by *LCY*.
 Used to compare suitability of candidate commas for prime commas. Lower is better.

*CY*  The *Complexity* function, which for a reduced fraction $x/y$ equals $xy$.
 Also called 'Benedetti Height'.

$C_z$  An octave shift; a 2–limit JI frequency, with reference point $C_4 = 1/1$.

free–JI  Just Intonation with specifically no limit on which primes can be used

Identity  The identity notation is $C_4 = 1/1$, which is the trivial notation / trivial component.

*LCY*  The *LogComplexity* function, equal to $\log_2 CY$ which is the 'Tenney Height'

| | |
|---|---|
| $L_z$ | Notation for a Pythagorean (3–limit) JI note. L is the note label, $z$ is the octave number. |
| L[$x/y$] | Notation for any free–JI pitch class |
| L[$x/y$]$_z$ | Notation for any free–JI frequency which can be any rational number |
| JI | Just Intonation, a system where relative frequencies form whole–numbered ratios |
| $N_4$ | A Pythagorean diatonic scale note from ($C_4$, $D_4$, $E_4$, $F_4$, $G_4$, $A_4$, $B_4$) |
| Octave | A frequency ratio of 2/1 |
| [$p$] | Prime comma for prime $p$; a fraction which is close to 1/1 |
| $p$–limit | Just Intonation using only $p$–smooth frequencies |
| $p$–rough | Rational numbers with all prime factors greater than or equal to $p$ |
| $p$–smooth | Rational numbers with all prime factors less than or equal to $p$ |
| Pythagorean | JI frequencies of the form $2^a 3^b$, i.e. 3–limit frequencies, rational numbers having only prime factors 2 and 3 |
| Semitone | A (non–JI) interval one–twelfth the size of an octave, of size 100 cents or $2^{1/12} = 1.0595\ldots$ |
| Simple | A simple notation requires exactly 1 non–trivial component to describe it |
| # | Sharp character, equivalent to fraction 2187/2048 |
| $S_k$ | A 'sharp number' representing one or more sharps '#' or flats 'b' or none ($S_0 = 1/1$) |
| Trivial | A trivial notation component is equal to $C_4 = 1/1$ |
| [$x/y$] | Rational comma built up by multiplying or dividing prime commas |

Shorthand notation

| | |
|---|---|
| ' | Prime character, shorthand for prime comma [5] = 80/81<br>e.g. E'$_4$ is the same as E[5]$_4$ |
| . | Period character, shorthand for rational comma [1/5] = 81/80<br>e.g. Ab.$_4$ is the same as Ab[1/5]$_4$ |
| ~ | Tilde character, provides shorthand L~$p$ for pitch class L[$p$].<br>Not recommended for notating specific frequencies, due to confusion with placement of octave number. |
| _ | Underscore character, provides shorthand L_$p$ for pitch class L[1/$p$].<br>Not recommended for notating specific frequencies. |

## 14) Author contact details

| | | |
|---|---|---|
| ORC ID | http://orcid.org/0000-0002-4785-9766 | Academic profile |
| arXiv | https://arxiv.org/a/ryan_d_1 | Papers (pre-prints) |
| SoundCloud | https://soundcloud.com/daveryan23/tracks | JI music examples |
| LinkedIn | https://www.linkedin.com/in/davidryan59 | Professional page |
| Email: | david ryan 1998 @ hotmail.com | (remove spaces) |